\documentclass[11pt]{article}
\usepackage{moriond,epsfig}

\def\be{\begin{equation}}
\def\ee{\end{equation}}
\def\bea{\begin{eqnarray}}
\def\eea{\end{eqnarray}}

\def\lsim{\raise0.3ex\hbox{$<$\kern-0.75em\raise-1.1ex\hbox{$\sim$}}}
\def\gsim{\raise0.3ex\hbox{$>$\kern-0.75em\raise-1.1ex\hbox{$\sim$}}}

\def\beq{\begin{equation}}
\def\eeq{\end{equation}}
\def\bea{\begin{eqnarray}}
\def\eea{\end{eqnarray}}
\def\bq{\begin{quote}}
\def\eq{\end{quote}}

\newcommand{\ie}{{\it i.e.}}

\def\lsim{\raise0.3ex\hbox{$<$\kern-0.75em\raise-1.1ex\hbox{$\sim$}}}
\def\gsim{\raise0.3ex\hbox{$>$\kern-0.75em\raise-1.1ex\hbox{$\sim$}}}

\def\dAu  {$d$Au}

\def\AuAu {AuAu}
\def\pp   {$pp$}
\def\pA   {$pA$}
\def\AA   {$AA$}
\def\AB   {$AB$}

\def\sqrtsNN {\mbox{$\sqrt{s_{NN}}$}}
\def\Npart   {\mbox{$N_{\rm part}$}}
\def\Ncoll   {\mbox{$N_{\rm coll}$}}

\def\RdAu    {\mbox{$R_{d\rm Au}$}}

\def\jpsi    {\mbox{$J/\psi$}}
\def\pT      {\mbox{$P_{T}$}}
\def\beq     {\begin{equation}}
\def\eeq     {\end{equation}}

\long\def\symbolfootnote[#1]#2{\begingroup%
  \def\thefootnote{\fnsymbol{footnote}}\footnote[#1]{#2}\endgroup}

\begin{document}
\title{COLD NUCLEAR MATTER EFFECTS ON \jpsi\ PRODUCTION AT RHIC: COMPARING SHADOWING MODELS}

\author{
E. G. FERREIRO$^1$, F. FLEURET$^2$, J. P. LANSBERG$^3$, A. RAKOTOZAFINDRABE$^4$}

\address{
$^1$Departamento de F\'{\i}sica de Part\'{\i}culas,
Universidade de Santiago de Compostela \\
E-15782 Santiago de Compostela, Spain}
\address{$^2$Laboratoire Leprince Ringuet, CNRS-IN2P3, \'Ecole Polytechnique,   91128 Palaiseau, France}
\address{$^3$SLAC National Accelerator Laboratory, Theoretical Physics, Stanford University,\\
Menlo Park, CA 95025, USA
}
\address{$^4$IRFU/SPhN, CEA Saclay, 91191 Gif-sur-Yvette Cedex, France}

\maketitle\abstracts{
We present a wide study on the comparison of different shadowing models and
their influence on \jpsi\ production. We have taken into account the
possibility of different partonic processes for the
$c \bar c$-pair production. We notice that the effect of shadowing corrections
on \jpsi\ production clearly depends on the partonic process considered.
Our results are compared to the available data on \dAu\ collisions
at RHIC energies. We try different break up cross section for each of the studied shadowing
models.
}


\section{Introduction}
\label{sec:intro}

The interest devoted to \jpsi\ production has not decreased for the last thirty years.
It is motivated by the search of the transition
from hadronic matter to a deconfined state of QCD matter,
the so-called Quark-Gluon Plasma (QGP).
The high density of gluons in the QGP is expected to hinder the formation of quarkonium systems,
by a process analogous to Debye
screening of the electromagnetic field in a plasma \cite{Matsui86}.

The results on \jpsi\ production measured by the PHENIX experiment at the
BNL Relativistic Heavy Ion Collider (RHIC) show
a significant suppression of the \jpsi\ yield in \AuAu\ collisions at \sqrtsNN=200~GeV \cite{Adare:2006ns}.
Nevertheless, PHENIX data on \dAu\ collisions \cite{Adare:2007gn} have also
pointed out that Cold Nuclear Matter (CNM) effects play an essential
role at these energies.

All this reveals that, in fact,
the interpretation of the results obtained in nucleus-nucleus collisions relies on a good
understanding and a proper subtraction of the CNM effects, known to
impact the \jpsi\ production in proton(deuteron)-nucleus collisions, where the
deconfinement can not be reached.

It is our purpose here to develop an exhaustive study of these effects.
This includes
shadowing, \ie~the modification of the parton distribution of a nucleon in a nucleus,
and final-state nuclear absorption.

Moreover, we have recently noticed \cite{OurExtrinsicPaper} that the impact of gluon shadowing
on \jpsi\ production does depend on the partonic
process producing the $c \bar c$ and then the \jpsi. Because of this, we have included here
two different production mechanism:  $g+g \to c\bar c \to J/\psi \,(+X)$, that corresponds to the
picture of the Colour Evaporation Model (CEM) at LO,
and $g+g \to J/\psi +g$, as in the Color Singlet Model (CSM), but including the $s$-channel cut contributions.

In practice, we have proceeded as follows:
we have developed a Glauber Monte-Carlo code where we have
interfaced the two different partonic processes for $c \bar c$ with the CNM effects, namely
the shadowing and nuclear absorption, in order to get the \jpsi\ production cross sections
for \pA\ and \AA\ collisions. We have considered four different models for the shadowing effects.
We shall compare our results with the experimental
measurements on \dAu\ collisions presently available at RHIC.


\section{The Model}
\label{sec:ourapproach}


To describe the \jpsi\ production in nucleus collisions, our Monte-Carlo
framework is based on the probabilistic Glauber model, the nuclear density
 profiles being defined with the Woods-Saxon parameterisation for
any nucleus ${A>2}$ and the Hulthen wavefunction for the
deuteron.
The nucleon-nucleon inelastic cross section at
$\sqrtsNN=200\mathrm{~GeV}$ is taken to $\sigma_{NN}=42\mathrm{~mb}$ and the
average nucleon density to $\rho_0=0.17\mathrm{~nucleons/fm}^3$. For each event
(for each \AB~collision) at a random impact parameter~$b$, the Glauber Monte-Carlo
model allows us to determine the number of nucleons in the path of each incoming
nucleon, therefore allowing us to easily derive the number \Ncoll\ of
nucleon-nucleon collisions and the total number \Npart\ of nucleons
participating into the collision.
In order to study the \jpsi\ production, we need to implement
in our Monte Carlo the following ingredients: the partonic process for the
$c \bar c$ production and the CNM effects.

\subsection{Partonic process for the $c \bar c$ production}

Most of the
studies of \jpsi\ production rely on the assumption that the $c \bar c$  pair is produced by the fusion of two gluons
 carrying some intrinsic transverse momentum~$k_T$. The partonic process being a $2\to 1$ scattering,
the sum of the gluon intrinsic transverse momentum is transferred to the $c \bar c$ pair, thus to
the \jpsi\ since the soft hadronisation process does not modify the kinematics. This corresponds
to the picture of the Colour Evaporation Model (CEM) at LO.
In such approaches, the transverse momentum of the \jpsi\ {\it entirely} comes from the intrinsic
transverse momentum of the initial gluons.

However, such an effect is not sufficient to describe the $P_T$ spectrum of quarkonia produced in
hadron collisions~\cite{Lansberg:2006dh}. Most of the transverse momentum should have an extrinsic
origin, \ie\ the \jpsi's $P_T$ would be balanced by the emission of a recoiling particle in the final
state. 
The \jpsi\ would then be produced by gluon fusion in a $2\to 2$ process with emission of a hard final-state gluon.
This emission, which is anyhow mandatory to conserve $C$-parity, 
has a definite influence on the kinematics of the
\jpsi\ production. Indeed, for a given \jpsi\ momentum (thus for
fixed~$y$ and $P_T$), the processes discussed above, \ie\ $g+g \to c\bar c \to J/\psi \,(+X)$
and $g+g \to J/\psi +g$,  will proceed on the average from gluons with different Bjorken-$x$. Therefore,
 they will be affected by different shadowing corrections. From now on, we will refer to the former
scenario as the {\it intrinsic} scheme, and to the latter as the {\it extrinsic} scheme.

In the intrinsic scheme, the initial
gluons carry a non-zero intrinsic transverse momentum, which is transferred to the \jpsi\ .
Note that, following~\cite{OurIntrinsicPaper}, we do not neglect the value of the
 \jpsi's $P_T$ in this simplified kinematics.
In fact, in this scheme, we use the fits to the $y$ and \pT\ spectra measured
by PHENIX~\cite{Adare:2006kf} in \pp\ collisions at $\sqrt{s_{NN}}=200\mathrm{~GeV}$
as inputs of the Monte-Carlo.
The
measurement of the \jpsi\ momentum completely fixes the longitudinal
 momentum fraction carried by the initial partons:
\begin{equation}
x_{1,2} = \frac{m_T}{\sqrt{s_{NN}}} \exp{(\pm y)} \equiv x_{1,2}^0(y,P_T),
\label{eq:intr-x1-x2-expr}
\end{equation}
with the transverse mass $m_T=\sqrt{M^2+P_T^2}$, $M$ being the \jpsi\ mass.

On the other hand, in the extrinsic scheme, information from the data alone
-- the $y$ and \pT\ spectra -- is not sufficient to determine $x_1$ and $x_2$.
Indeed, the presence of a final-state gluon authorizes much more freedom to
choose $(x_1, x_2)$ for a given set $(y, P_T)$.
 The four-momentum conservation explicitely results in a more complex expression of $x_2$ as a function of~$(x_1,y,P_T)$:
\begin{equation}
x_2 = \frac{ x_1 m_T \sqrt{s_{NN}} e^{-y}-M^2 }
{ \sqrt{s_{NN}} ( \sqrt{s_{NN}}x_1 - m_T e^{y})} \ .
\label{eq:x2-extrinsic}
\end{equation}
Equivalently, a similar expression can be written for $x_1$ as a function of~$(x_2,y,P_T)$.
Even if the kinematics determines
the physical phase space, models are anyhow mandatory to compute the proper
weighting of each kinematically allowed $(x_1, x_2)$. This weight is simply
the differential cross section at the partonic level times the gluon Parton Distribution
Functions (PDFs),
\ie\ $g(x_1,\mu_f) g(x_2, \mu_f) \, d\sigma_{gg\to J/\psi + g} /dy \, dP_T\, dx_1 dx_2 $.
In the present implementation of our code, we are able to use the partonic differential
cross section computed from {\it any} theoretical approach. For now, we use the one
from~\cite{Haberzettl:2007kj} which takes into account the $s$-channel cut
contributions~\cite{Lansberg:2005pc} to the basic Color Singlet Model (CSM) and
satisfactorily describes the data down to very low~\pT~\ .

\subsection{Shadowing}

To get the \jpsi\ yield in \pA\ and \AA\ collisions, a shadowing-correction
factor has to be applied to the \jpsi\ yield obtained from the simple
superposition of the equivalent number of \pp\ collisions.
This shadowing factor can be expressed in terms of the ratios $R_i^A$ of the
nuclear Parton Distribution Functions (nPDF) in a nucleon of a nucleus~$A$ to the
PDF in the free nucleon. We will consider three
different shadowing models for comparation: EKS98~\cite{Eskola:1998df}, EPS08~\cite{Eskola:2008ca} and 
nDSg~\cite{deFlorian:2003qf} at LO.

These models
provide the nuclear ratios $R_i^A$ at a given initial value of $Q^2_0$ which is assumed
large enough for perturbative evolution to be
applied:
$Q^2_0 = 2.25\mathrm{~GeV^2}$ for EKS98,
$Q^2_0 = 1.69\mathrm{~GeV^2}$ for EPS08
and  $Q^2_0 = 0.4\mathrm{~GeV^2}(0.26\mathrm{~GeV^2}$) for nDSg
and perform their evolution through the DGLAP evolution
equations to LO accuracy in the case of EKS98 and EPS08 and to LO and NLO accuracy in the case of nDSg.
The spatial dependence of the shadowing is not given in the above models. However,
it has been included in our approach, assuming that the inhomogeneous shadowing is proportional to the local
density~\cite{Klein:2003dj}.

The nuclear ratios of the PDFs are then expressed by:
\beq
\label{eq4}
R^A_i (x,Q^2) = \frac{f^A_i (x,Q^2)}{ A f^{nucleon}_i (x,Q^2)}\ , \ \
f_i = q, \bar{q}, g \ .
\eeq
The numerical parameterisation of $R_i^A(x,Q^2)$
is given for all parton flavours. Here, we restrain our study to gluons since, at
high energy, \jpsi\ is essentially produced through gluon fusion \cite{Lansberg:2006dh}.

\subsection{The nuclear absorption}

The second CNM effect that we are going to take into account concerns
the nuclear absorption.  In the framework of the probabilistic Glauber
model, this effect refers to the probability for the pre-resonant $c{\bar c}$
pair to survive to the propagation through the nuclear medium and is usually parametrised
by introducing an effective absorption cross section~$\sigma_{\mathrm{abs}}$.
It is our purpose here to compare different absorptive $\sigma$ 
within the two partonic $c \bar c$ production mechanisms --intrinsic and extrinsic-- and for the three shadowing models cited above.

\section{Results}

In the following, we present our results for the  \jpsi\ nuclear modification factor:
\beq
R_{AB}=\frac{dN_{AB}^{J/\psi}}{\langle\Ncoll\rangle dN_{pp}^{J/\psi}}.
\eeq
$dN_{AB}^{J/\psi} (dN_{pp}^{J/\psi})$ is the \jpsi\ yield observed in \AB\ (\pp) collisions
and $\langle\Ncoll\rangle$ is the average number of nucleon-nucleon collisions occurring
in one \AB\ collision. In the absence of  nuclear effects, $R_{AB}$ should equal unity.

We will restrict ourselves to  \dAu\ collisions,
since only CNM matter effects are at play here, so they
provide the best field for the study of the shadowing and the nuclear absorption.

\begin{figure*}[htb!]
\begin{center}
\includegraphics[width=1.0\linewidth]{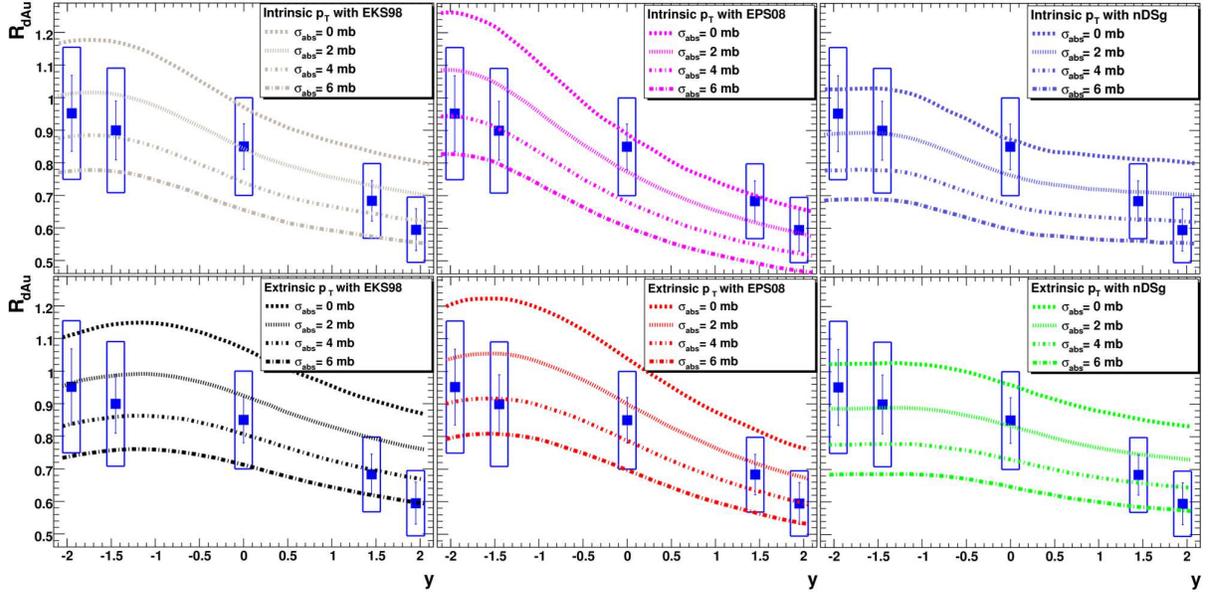}
\end{center}
\caption{\jpsi\ nuclear modification factor in \dAu\ collisions at $\sqrt{s_{NN}}=200\mathrm{~GeV}$ versus $y$ for the
different shadowing models in the intrinsic (up) and extrinsic (down) scheme.
For complementary figures see: \tt http://phenix-france.in2p3.fr/software/jin/index.html}
\label{fig:RdAu_vs_y}
\end{figure*}

We have used PHENIX measurements of \RdAu~\cite{Adare:2007gn} in order 
to compare the different shadowing models.
In Fig.~\ref{fig:RdAu_vs_y}, we have computed our results in the different shadowing frameworks for  
four $\sigma_{\mathrm{abs}}$ for each of the shadowing models considered
in both the intrinsic and extrinsic scheme.

We have also evalute the best fit for $\sigma_{\mathrm{abs}}$,
following the method used by PHENIX in \cite{Adare:2007gn} and \cite{Adare:2008cg}.
By using the data on  \RdAu\ versus rapidity, we have obtained the best $\chi^2$
for the EPS08 model, computed in the extrinsic scheme for a  $\sigma_{\mathrm{abs}}=3.6$ mb.

\vskip -0.5cm
\section*{Acknowledgments}
E. G. F. thanks Xunta de Galicia (2008/012) and 
Ministerio de Educacion y Ciencia of Spain (FPA2008-03961-E/IN2P3)
for financial support.


\end{document}